\documentclass[twocolumn,preprintnumbers,floatfix,prb,superscriptaddress,aps]{revtex4-2}
\usepackage{graphicx}
\usepackage{float}
\usepackage{xcolor}
\usepackage{dcolumn} 
\usepackage{bm}
\usepackage{amsmath}
\usepackage{makecell}
\usepackage{multirow}
\usepackage{hyperref}
\usepackage{subfigure}
\hypersetup{colorlinks=true,
	    final=true,
	    linkcolor=blue,
	    citecolor=blue,
	    filecolor=blue,
	    urlcolor=blue,
	    }
\newcommand{\edit}[1]{{\textbf{#1}}}

\begin{document} 

\title{\edit{Antiferromagnetic insulating state in layered nickelates at half-filling}}

\author{Myung-Chul Jung}
\affiliation{Department of Physics, Arizona State University, Tempe, AZ 85287, USA}
\author{Harrison LaBollita}
\affiliation{Department of Physics, Arizona State University, Tempe, AZ 85287, USA}

\author{Victor Pardo}
 \email{victor.pardo@usc.es}
\affiliation{Instituto de Materiais iMATUS, Universidade
de Santiago de Compostela, E-15782 Santiago de Compostela,
Spain
}
\affiliation{
Departamento de F\'{\i}sica Aplicada, Universidade
de Santiago de Compostela, E-15782 Santiago de Compostela,
Spain
}

\author{Antia S. Botana}
\email{Antia.Botana@asu.edu}
\affiliation{Department of Physics, Arizona State University, Tempe, AZ 85287, USA}


\begin{abstract}

We provide a set of computational experiments based on \textit{ab initio} calculations to elucidate whether a cuprate-like antiferromagnetic insulating state can be present in the phase diagram of the low-valence layered nickelate family (R$_{n+1}$Ni$_n$O$_{2n+2}$, R= rare-earth, $n=1-\infty$) in proximity to half-filling. It is well established that at $d^9$ filling the infinite-layer ($n=\infty$) nickelate is metallic, in contrast to cuprates wherein an antiferromagnetic insulator is expected. We show that for the Ruddlesden-Popper (RP) reduced phases of the series (finite $n$) an antiferromagnetic insulating ground state can naturally be obtained instead at $d^9$ filling, due to the spacer RO$_2$ fluorite slabs present in their structure that block the $c$-axis dispersion. In the $n=\infty$ nickelate, the same type of solution can be derived if the off-plane R-Ni coupling is suppressed. 
We show how this can be achieved if a structural element that cuts off the $c$-axis dispersion is introduced (i.e. vacuum in a monolayer of RNiO$_2$, or a blocking layer in multilayers formed by (RNiO$_2$)$_1$/(RNaO$_2$)$_1$). 
\end{abstract}

\maketitle

\section{Background}
\begin{figure}[h!]
    \centering
    \includegraphics[width=\columnwidth]{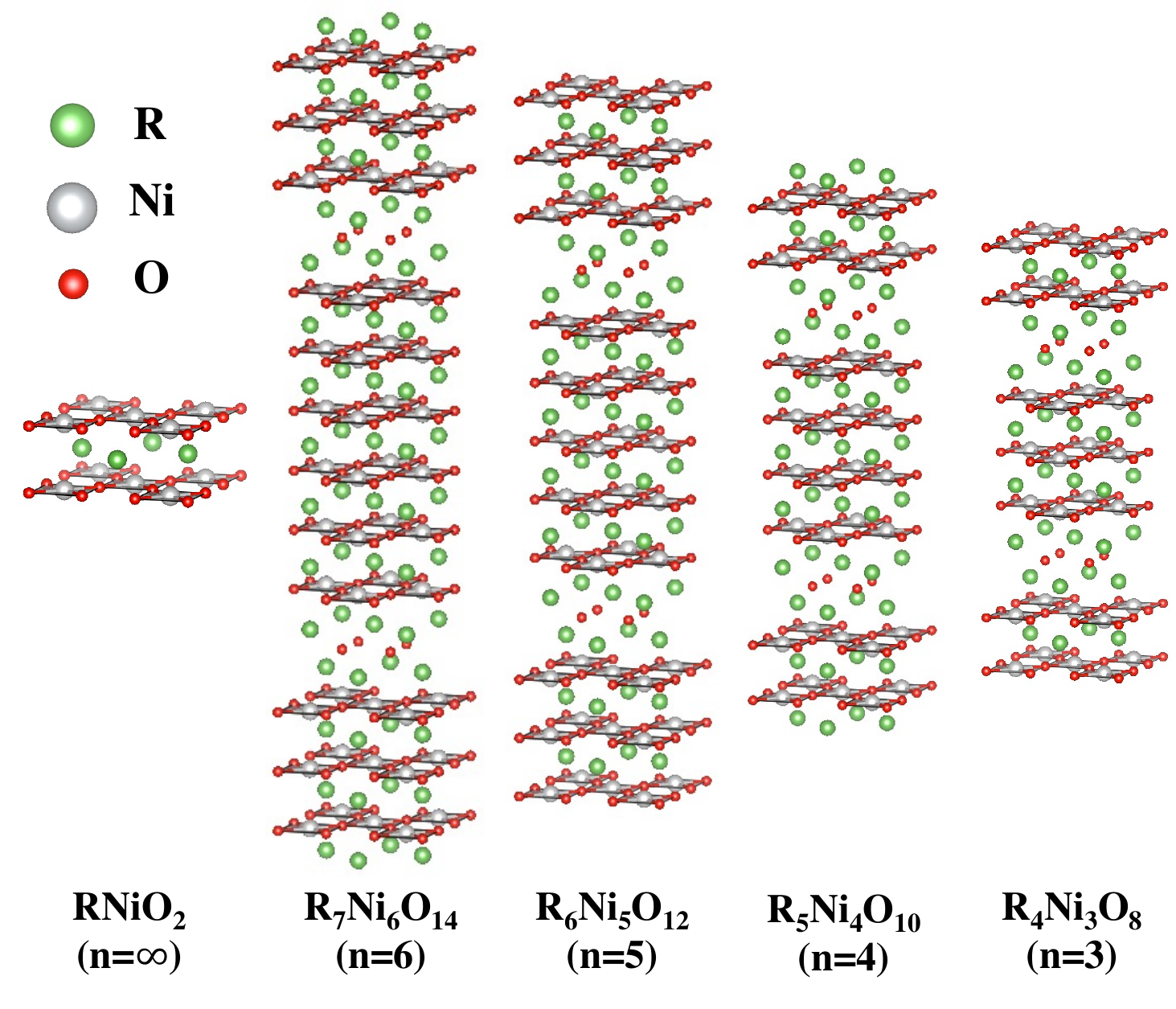}
    \caption{Structure of the oxygen-reduced  R$_{n+1}$Ni$_n$O$_{2n+2}$ series for $n=3-6$ and $n=\infty$. The common structural motifs can be observed: NiO$_2$ planes with a square-planar coordination of oxygens around the Ni cations. Fluorite RO$_2$ blocking slabs that separate series of $n$-NiO$_2$ layers appear for $n=3-6$.}
    \label{fig1}
\end{figure}

Superconductivity in cuprate-like systems has been long sought for \cite{norman-RPP}. Among different approaches to achieve cuprate-analog materials, targeting nickelates has been an obvious strategy as nickel and copper are next to each other in the periodic table \cite{anisimov_lanio2}. After a 30-year search, Li and coworkers reported superconductivity in Sr-doped infinite-layer NdNiO$_2$ in 2019 \cite{hwang_ndnio2_sc, sc_dome_nickelates}, and more recently in PrNiO$_2$  \cite{osada2020prnio2sc, prnio2_2020_phase_diagram} and LaNiO$_2$ \cite{zeng2021superconductivity}. These infinite-layer nickelates are formed by NiO$_2$ planes (with a square-planar oxygen environment for the cation, like the CuO$_2$ planes in cuprates), with Ni adopting the unusual Ni$^{1+}$ valence, with nine $d$-electrons, (analogous to Cu$^{2+}$) \cite{pickett_review_cuprates,botana_2020_similarities}. 
As of now, superconductivity has only been observed in thin films, prompting work on the role played by interfacial effects \cite{dagotto, pentcheva}.
Importantly, RNiO$_2$ (R= rare-earth) materials belong to a larger series represented by R$_{n+1}$Ni$_{n}$O$_{2n+2}$ ($n = 1- \infty$) in which each member contains $n$-NiO$_{2}$ layers \cite{greenblatt1997ruddlesden} opening up the door to find a whole family of nickelate superconductors (see Fig. \ref{fig1}). Indeed, very recently the $n$= 5 Nd-based member of the series (Nd$_6$Ni$_5$O$_{12}$) was found to be a superconductor as well \cite{pan2021superconductivity}, and proposals exist for superconductivity to be realized in the $n$= 3 compound (R$_4$Ni$_3$O$_8$) upon electron doping \cite{physrevmat}, so that it falls within the superconducting dome in terms of $d$-filling in a general phase diagram applicable for the whole series. 

In spite of the similarities between this nickelate family and the cuprates  described above, a remarkable difference is that while parent cuprates are antiferromagnetic (AF) insulators \cite{pickett_review_cuprates}, parent infinite-layer nickelates are metallic and there is no signature of long-range magnetic order in LaNiO$_2$ \cite{lanio2_structure_2009,lanio2_structure_2005,lanio2_structure_2016} or NdNiO$_2$  \cite{ndnio2_structure_2003}.  
Based on these findings, magnetism was at first ruled out as being a key ingredient for superconductivity in RNiO$_2$-based  systems, even though many theories of superconductivity in cuprates rely on the existence of strong AF correlations \cite{anderson1987resonating}. Recent RIXS experiments in NdNiO$_2$ point in this direction as they have revealed the presence of strong AF correlations with a superexchange $\sim$ 60 meV \cite{Lu2021_NdRIXS}. Also, an NMR study has found the presence of AF fluctuations and quasi-static AF order below 40~K in Sr-doped NdNiO$_2$ \cite{nmr_2020_cui}.   
Along these lines, first-principles calculations do predict that the main correlations in RNiO$_2$ are indeed AF, but less strong than in the cuprates \cite{botana_2020_similarities,KWLee2004_prb,Choi2020,Liu2020,Gu2020}, and embedded in a metallic environment. While magnetism in these infinite-layer nickelates is still under intense study \cite{choi2020fluctuation, krishna2020effects, kapeghian2020electronic,nmr_2020_cui}, it is then important to understand: i) Why parent infinite-layer nickelates are metallic and do not show any signature of long-range magnetic order (their cuprate analog CaCuO$_2$ is AF and insulating \cite{cacuo2singh1989electronic}). ii) Whether the phase diagram of these newly discovered superconducting nickelates  can host an AF insulating phase close to half-filling by modifying parameters such as dimensionality. iii) If other (finite $n$) members of the layered nickelate family are more similar to cuprates at $d^9$ filling in terms of their electronic structure.

In this paper, using first-principles calculations, we try to address these questions. We argue that the metallic non-cuprate-like behavior of parent infinite-layer nickelates can be attributed to the strong Ni-R-Ni  off-plane hopping. This gives rise to extremely dispersive R-$d$ bands, strongly hybridized with the Ni-$d$ occupied bands. These R-$d$ bands cross the Fermi level leading to a deviation from half-filling in the Ni-$d_{x^2-y^2}$ band 
via a self doping mechanism. We show that if structural modifications are used to provide a means to suppress the Ni-Ni off-plane hopping, an AF insulating state can be obtained.
As such, we reveal that in the purely two-dimensional limit, RNiO$_2$ materials are indeed AF insulators. This very same phase can be realized naturally in the finite-$n$ members of the layered nickelate family (with a fluorite blocking slab between NiO$_2$ planes), including the recently found $n=5$ layered nickelate superconductor. As such, our calculations show that these finite-$n$ nickelates can be AF insulators at $d^9$ filling without any structural changes. All in all, our results illustrate that a crucial part of the phase diagram of the superconducting cuprates (the parent insulating antiferromagnetic phase) can indeed be present for the whole series of layered nickelates. In the case of the $n$ = $\infty$ compound it is simply hindered by strong off-plane band dispersions but a dimensionality reduction can bring it to light.

\section{Computational methods}

Our electronic structure calculations in layered nickelates were performed within density functional
theory \cite{dft} using the all-electron, full potential code {\sc wien2k} \cite{wien2k1,wien2k2}
based on the augmented plane wave plus local orbital (APW+lo) basis set \cite{sjo} and the Vienna $ab$ $\it{initio}$ Simulation Package (VASP)\cite{vasp1}.  We present the results obtained for R=La in all materials to avoid problems associated with the $4f$ electrons of Pr or Nd but no large changes can be anticipated from the R substitution in the electronic structure, as shown for the RNiO$_2$ materials in a number of works~\cite{Uchida2011_prl,botana_2020_similarities,Ryee2020_prb,kapeghian2020electronic,Been2021_prx}. We did indeed perform calculations in RNiO$_2$ for a different rare-earth (Nd) and only minor changes not affecting the main conclusions of the paper arose, which will be explained in the Appendix.
The generalized gradient approximation in the Perdew-Burke-Ernzerhof (GGA-PBE) scheme \cite{gga} was used as the exchange-correlation functional.
To deal with possible strong correlation effects we use GGA+$U$ \cite{sic} including an on-site repulsion $U$ ($U$ = 5 eV) and Hund's coupling $J$ ($J$= 0.7 eV) for the Ni $3d$ states.

For LaNiO$_2$, when thin films are discussed, a sufficient vacuum of 20 \AA\  was introduced, enough to guarantee the lack of interaction between two planes (periodic boundary conditions are utilized along the three spatial directions). Both in thin films and multilayers, atomic positions were fully relaxed within GGA-PBE using the in-plane lattice parameter of LaNiO$_2$. In both cases, relaxations were done in a checkerboard antiferromagnetic state in a $\sqrt{2}$$\times$$\sqrt{2}$ supercell of the tetragonal ($P4/mmm$) cell. A C-type antiferromagnetic state has been chosen in the bulk as this is the ground state of the system as soon as a small $U$ is included \cite{kapeghian2020electronic}.  The muffin-tin radii used were: 2.50 a.u. for Nd and La, 2.00 a.u. for Ni and 1.72 a.u. for O. An $R_{mt}K_{max}$ = 7.0 was chosen for all calculations. 

For the La-based $n=4-6$ compounds  (La$_5$Ni$_4$O$_{10}$ ($n=4$), La$_6$Ni$_5$O$_{12}$ ($n=5$), and La$_7$Ni$_6$O$_{14}$ ($n=6$)), whose structures have not been experimentally resolved yet, we conducted structural relaxations using VASP within GGA-PBE, optimizing both lattice parameters and internal coordinates in a $\sqrt{2}$$\times$$\sqrt{2}$ supercell of the tetragonal ($I4/mmm$) cell in a checkerboard (C-type) antiferromagnetic state. As above, a C-type antiferromagnetic state has been chosen as this is the ground state of the finite $n$ materials as soon as a small $U$ is included to account for correlations in the Ni 3$d$ states.
In the VASP calculations, an energy cutoff of 650 eV and a $k$-mesh of 6$\times$6$\times$1 were adopted with a force convergence criterion of 5 meV/\AA. 

With the VASP-optimized structures for the $n=4-6$ nickelates, we then performed further electronic structure calculations using the {\sc wien2k} code. In these calculations, muffin-tin radii of 2.50 a.u., 2.00 a.u., and 1.72 a.u. were used for La, Ni, and O, respectively. We used $R_{mt}K_{max}$ = 7.0 and a $k$-grid of 21$\times$21$\times$21 for all materials in the irreducible Brillouin zone.
To modify the chemical potential in the finite-$n$ compounds, the virtual crystal approximation (VCA) was used in both the La-site and Ni-site (to achieve a nominal $d^9$ filling for all materials), both yielding consistent results.

\begin{table}[h!]
\caption{Lattice parameters and bond lengths obtained from the structural relaxations for $n=4-6$ layered nickelates. The experimental structural data for the $n=3$ compound \cite{Junjie2016_pnas} are shown as a reference. Labels $i$, $m$, and $o$ denote the inner-, mid-, and outer-layers, respectively.}
\begin{center}
\begin{tabular*}{\columnwidth}{l@{\extracolsep{\fill}}cccccccc}
   \hline\hline
    &\multicolumn{8}{c}{higher-order layered nickelates}\\
            &~& \makecell{$n$=3 \\ (exp)} &~& $n$=4  &~& $n$=5 &~& $n$=6\\\cline{2-9}
    d (\AA) &~& \makecell{$a$=5.61  \\ $c$=26.09 } 
            &~& \makecell{$a$=5.62  \\ $c$=33.36 } 
            &~& \makecell{$a$=5.62  \\ $c$=40.20 } 
            &~& \makecell{$a$=5.62  \\ $c$=47.01 } \\\hline
    Ni($i$)-O       &~& 1.985  &~& 1.986  &~& 1.985  &~& 1.985 \\
    Ni($m$)-O       &~&        &~&        &~& 1.985  &~& 1.985 \\
    Ni($o$)-O       &~& 1.985  &~& 1.988  &~& 1.988  &~& 1.988 \\
    Ni($i$)-Ni($i$) &~&        &~& 3.385  &~&        &~& 3.378 \\
    Ni($i$)-Ni($m$) &~&        &~&        &~& 3.375  &~& 3.385 \\
    Ni($m$)-Ni($o$) &~&        &~&        &~& 3.281  &~& 3.286 \\
    Ni($i$)-Ni($o$) &~& 3.262  &~& 3.260  &~&        &~&       \\\hline\hline
\end{tabular*}
\end{center}
\label{str_data}
\end{table}

\section{Higher-order layered nickelates}
\label{section2}

In order to unravel the nature of the ground state of the whole layered-nickelate family when approaching the $d^9$ nominal filling, we will start by analyzing the electronic structure of the reduced RP compounds as $n$ is varied. As mentioned above, the $n=5$ Nd-based layered nickelate has been recently found to be superconducting, increasing the interest in these higher-$n$ layered nickelates. 

We have used the $I4/mmm$  structure of the $n= 3$ compound \cite{Poltavets2007_Ln4Ni3O8,cheng2012pressure} as a model for constructing the structures of the $n=4-6$ counterparts, since the latter have not been experimentally resolved yet. The structure of the $n=3-6$ materials (reduced from Ruddlesden-Popper phases) can be visually inspected in Fig. \ref{fig1}. 
The structure is formed by a series of $n$-NiO$_2$ planes where the Ni coordination is a square plane of oxygens. These $n$ planes are sandwiched by RO$_2$ fluorite blocking layers that confine the Ni-$d$ states along the $c$-direction conferring the Ni-$d_{z^2}$ bands a localized, molecular-like character \cite{pardo2010quantum}. The structural parameters (nearest-neighbor distances and lattice parameters) obtained after relaxation for the La-based $n=4-6$ nickelates are summarized in Table \ref{str_data} compared to the experimentally resolved ones for the $n=3$ material~\cite{Junjie2016_pnas}.

\begin{figure*}
\centering
    \includegraphics[width=\textwidth]{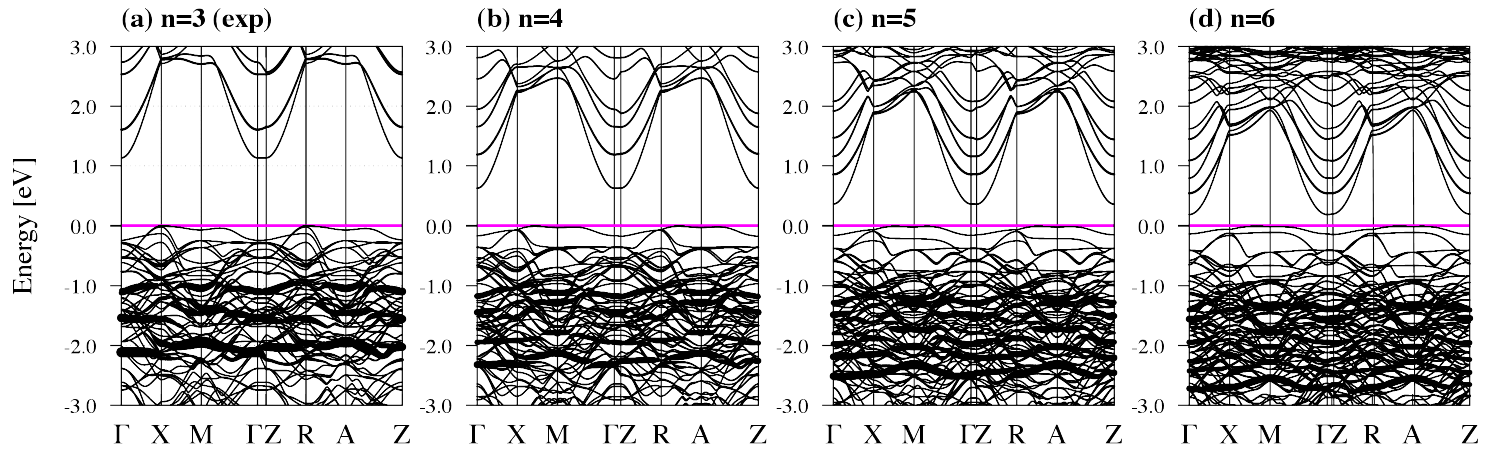}
    \caption{Band structures for La-based $n= 3-6$ reduced nickelates at $d^9$ filling using GGA+$U$ ($U$= 5 eV) with AF ordering (La$_4$Ni$_3$O$_8$= 438, $n=3$, La$_5$Ni$_4$O$_{10}$= 5410, $n=4$, La$_6$Ni$_5$O$_{12}$= 6512, $n=5$, and La$_7$Ni$_6$O$_{14}$= 7614, $n=6$). (a) For the $n= 3$ material, the energy gap is 1.13 eV. (b) For $n= 4$, the energy gap is 0.63 eV. (c) For $n= 5$, the energy gap is 0.35 eV. (d) For $n= 6$, the energy gap is 0.19 eV. The majority $d_{z^2}$ orbital of all Ni atoms is highlighted in the band structure.
    }
    \label{fig2}
\end{figure*}

Once the relaxations are carried out, we have artificially moved the chemical potential in all compounds (using the virtual crystal approximation) so that a nominal $d^9$ filling is achieved, in order to explore what magnetic and electronic phase is the ground state in this situation. The GGA checkerboard AF solutions at $d^9$ filling are metallic, a $U$ needs to be introduced on the Ni-$d$ states to open up a gap. A summary of the derived AF band structures within GGA+$U$ for the $n=3-6$ compounds is presented in Fig. \ref{fig2}. In all cases, we show the band structures for the ground state AF checkerboard C-type pattern, where the $S=1/2$ (nominally Ni$^{1+}$) cations couple AF due to the hole in the $d_{x^2-y^2}$ orbital. The derived Ni magnetic moments are $\sim$ 0.8-0.9 $\mu_B$ once a $U$ is added (with some slight variation in inner, outer, and mid-layers, as shown in Table \ref{mag_FLL_Ni}) and hence consistent with this Ni$^{1+}$ ($S=1/2$) picture. In all cases, a gap appears in the spectrum but its value gets reduced as $n$ increases, that is, as one moves towards the infinite-layer (three-dimensional) limit. This gap opens up between states that are predominantly $d_{z^2}$ in character at the top of the valence band and $d_{x^2-y^2}$ at the bottom of the conduction band, as expected (see Appendix \ref{appendix_A} Fig. \ref{fig6}). In this manner, the electronic structure resembles that of parent cuprates, where the ground state at $d^9$ filling is also an AF insulator. A similar result had been already reported for the $n= 3$ case, when the material is electron-doped with Ce, Zr, or Th \cite{physrevmat}.

These results give rise to the following questions: why are the infinite-layer counterparts metallic at the same nominal $d$-filling?, what happens to the AF correlations that should exist due to the partly filled $d_{x^2-y^2}$ band? In the following, we will try to answer this question carefully with the aid of electronic structure calculations through computational experiments in the infinite-layer materials.

\begin{table}[bt]
\caption{Ni magnetic moments for the $n=3-6$ nickelates within VCA at  $d^9$ filling within both GGA and GGA+$U$. Here, $i$, $m$, and $o$ denote the inner-, mid-, and outer-Ni atoms, respectively. }
\begin{center}
\begin{tabular}{c|c|cc}\hline\hline
    &~& \multicolumn{2}{c}{$d^9$ filling}\\\cline{3-4}
    compound & atom & GGA & \makecell{GGA+$U$} \\\hline
    \multirow{2}{*}{$n=3$} & Ni($i$) & $\pm$0.59 & $\pm$0.85 \\
        & Ni($o$) & $\pm$0.59 & $\pm$0.85 \\\hline
    \multirow{2}{*}{$n=4$} & Ni($i$) & $\pm$0.63 & $\pm$0.88 \\
        & Ni($o$) & $\pm$0.62 & $\pm$0.88 \\\hline
    \multirow{3}{*}{$n=5$} & Ni($i$) & $\pm$0.64 & $\pm$0.90 \\
        & Ni($m$) & $\pm$0.64 & $\pm$0.91 \\
        & Ni($o$) & $\pm$0.64 & $\pm$0.91 \\\hline
    \multirow{3}{*}{$n=6$} & Ni($i$) & $\pm$0.66 & $\pm$0.92 \\
        & Ni($m$) & $\pm$0.66 & $\pm$0.94 \\
        & Ni($o$) & $\pm$0.66 & $\pm$0.94 \\\hline\hline
\end{tabular}
\end{center}
\label{mag_FLL_Ni}
\end{table}

\section{Turning infinite layer nickelates into antiferromagnetic insulators.}
\label{section4}

\begin{figure}[ht]
\begin{center}
\includegraphics[width=\columnwidth,draft=false]{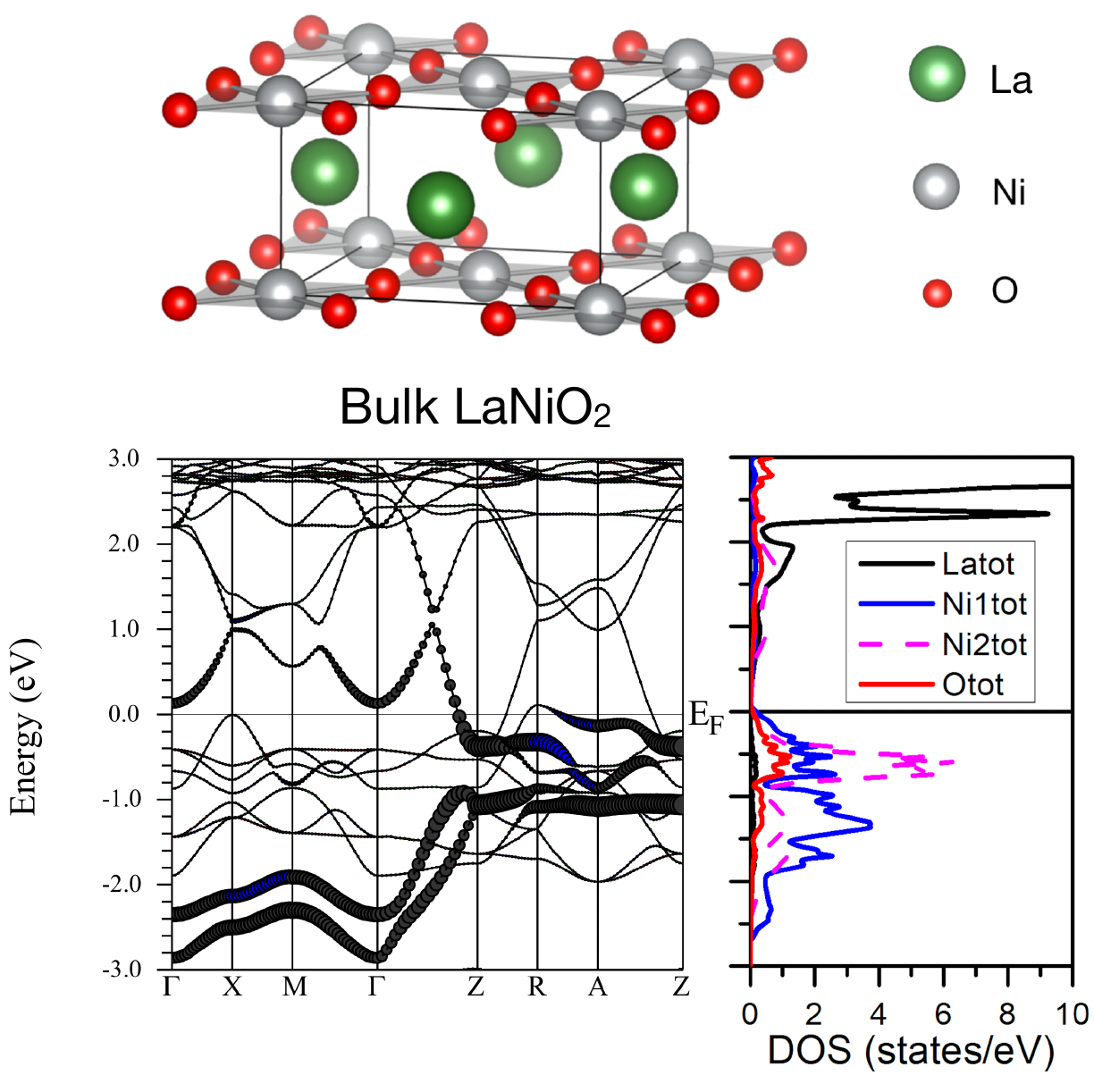}
\caption{Top panel. Structure of LaNiO$_2$. La atoms in green, Ni in grey, O in red. The structure highlights the NiO$_4$ square planar environment surrounding each Ni cation forming the NiO$_2$ building blocks of the layered structure in superconducting nickelates. Bottom left panel. Band structure of bulk LaNiO$_2$ within GGA in its AF ground state with the fat bands representing the Ni-$d_{z^2}$ states of both Ni atoms in the structure (a Hund splitting can be observed between them). Only one spin channel is shown, with the  very flat La-f bands appearing well above the Fermi level. The largely dispersive bands crossing the Fermi level are hybrid La-$d$ - Ni-$d$ bands and lead to metallic behavior in the bulk. Bottom right panel. Corresponding La, O, and Ni atom-resolved density of states  (both Ni atoms have opposite spins). }
\label{fig3}
\end{center}
\end{figure}

We start by describing the basic electronic structure of infinite-layer nickelates and how we can understand the differences between the $n= \infty$ case and the $n\neq \infty$ oxygen-reduced RP compounds described in the previous section.

Figure \ref{fig3} shows the GGA band structure and density of states (DOS) of bulk LaNiO$_2$ in its AF C-type (in-plane AF checkerboard) ground state. 
The metallic character can be seen to be brought about, as it has been analyzed in previous works \cite{choi2020fluctuation,kapeghian2020electronic}, by the presence of strongly dispersive bands (in particular along the $\Gamma$-Z direction) of mixed La-$d$ and Ni-$d_{z^2}$ character. 
In the non-magnetic state, La-$d$ bands also cross the Fermi level, as pointed out before \cite{botana_2020_similarities, Choi2020, Thomale_PRB2020}. Introducing a large $U$, either in the La-$d$ or the Ni-$d$ bands, does not lead to a band splitting that would open up a gap in this system \cite{choi2020fluctuation}. As such, the R-$d$  bands lead to a so-called self-doping effect, i.e. the depletion in occupation of the Ni-$d_{x^2-y^2}$ band, away from the nominal half-filling that would correspond to a Ni$^{1+}$ cation (see Appendix \ref{appendix_B} Fig. \ref{fig7} for the $d_{x^2-y^2}$ fat bands). Note that the presence of a flat Ni-$d_{z^2}$ band pinned at the Fermi level in connection to metallicity in the C-type AF state of infinite-layer nickelates has been highlighted in previous DFT work in connection to instabilities that
could limit AF order at low temperature \cite{choi2020fluctuation}.

\begin{figure}[ht]
\begin{center}
\includegraphics[width=\columnwidth,draft=false]{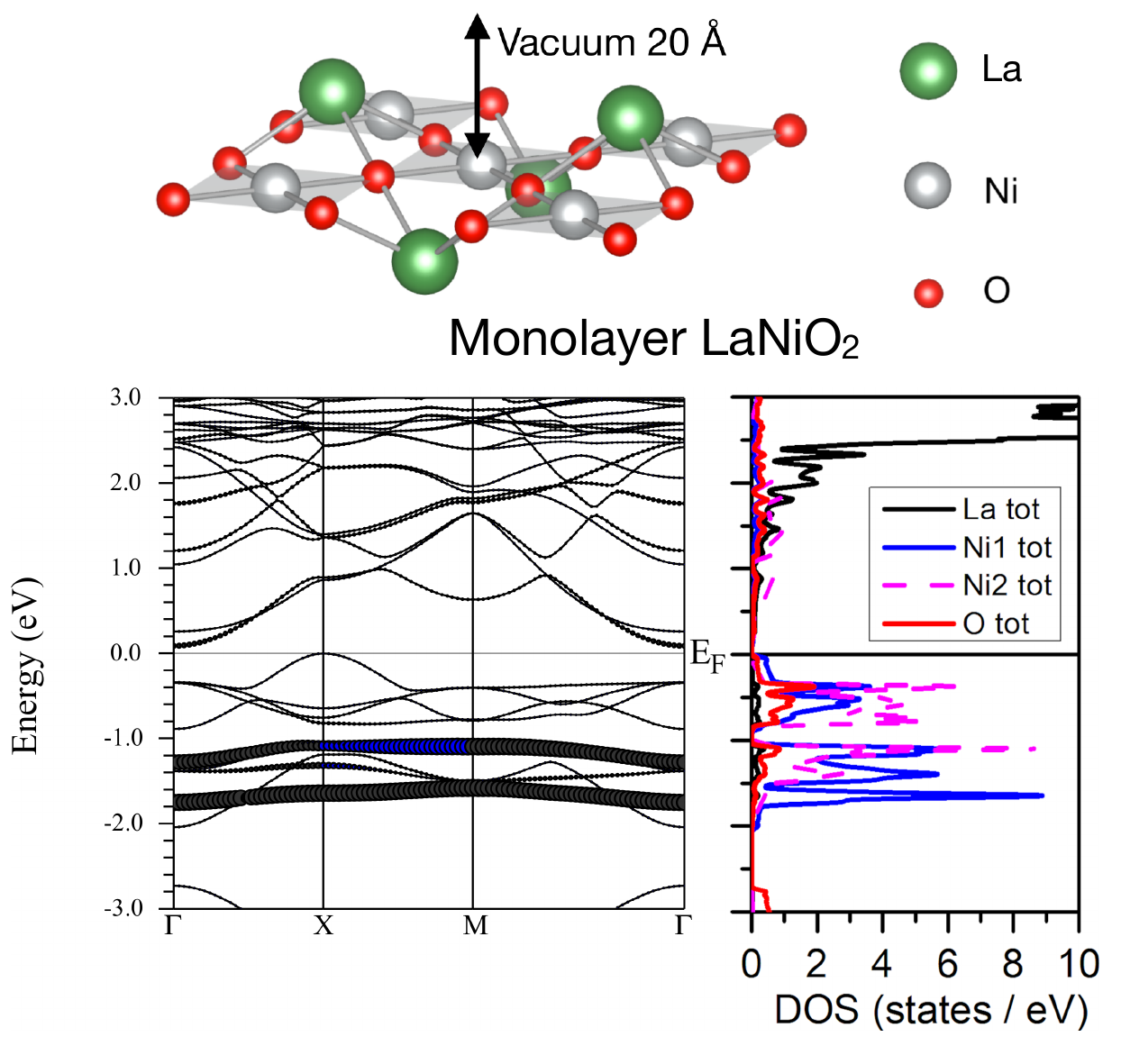}
\caption{Top panel. Structure of a  LaNiO$_2$  monolayer of the type used for the calculations discussed in the text. Bottom left panel. GGA band structure of the LaNiO$_2$ monolayer with AF ordering of the Ni atoms, showing only one spin channel. The Ni-$d_{z^2}$  states are highlighted for both Ni atoms (a Hund splitting can again be observed between them). The O-$p$ bands (not shown) start to appear below -3 eV. A gap opening  occurs. 
Bottom right panel. Corresponding La, O and  Ni atom-resolved density of states (both Ni atoms have opposite spins).}
\label{fig4}
\end{center}
\end{figure}

Now the question arises: is there a way one can get an insulating AF phase from here? To answer this question, we start by taking the system towards the purely two-dimensional limit. For that sake, we have  constructed an RNiO$_2$ monolayer from the constituting NiO$_2$ planes, with the standard square planar environment for the Ni cations (see Fig. \ref{fig4}). The additional rare-earth cations are placed alternatively on top and below the NiO$_2$ planes to ensure the stoichiometry and provide the highest possible symmetry. If the rare-earth cations are all placed on a single plane above or below the NiO$_2$ plane, the resulting total energy is much higher.

Figure \ref{fig4} shows the band structure of this single-layer LaNiO$_2$ system at the GGA level (without introducing correlations via  DFT+$U$) in an AF checkerboard ground state. 
The La-$d$ bands do not cross the Fermi level and the only Ni-$d$ band that remains unoccupied is a single minority-spin $d_{x^2-y^2}$ band per Ni (see Appendix \ref{appendix_B} Fig. \ref{fig8} for the $d_{x^2-y^2}$ fat bands). The La-$d$ bands cannot hybridize strongly with the Ni-$d$ ones anymore (once the periodicity along $c$ is lost) and as a consequence there is no largely dispersive band crossing the Fermi level, as it occurs in the bulk. As such, the system is no longer self-doped and the nominally Ni$^{1+}$ purely ionic description works nicely, like in parent cuprates and reduced RP nickelates. 
 The half-filled $d_{x^2-y^2}$ band yields a spin-1/2 scenario with a strong in-plane AF coupling, that appears even at the GGA level ($U$= 0 limit). This solution is similar to that obtained for cuprates (such as isostructural CaCuO$_2$) that are AF insulators at the GGA level as well \cite{botana_2020_similarities}. The magnetic moments within GGA are also somewhat larger ($\sim$ 0.9 $\mu_B$) than those appearing in bulk LaNiO$_2$ ($\sim$ 0.6 $\mu_B$) that get reduced by the self-doping effect. If an on-site Coulomb repulsion is introduced, the gap becomes widened as $U$ increases by pushing the unoccupied Ni-$d_{x^2-y^2}$ bands higher in energy, and the magnetic moments approach 1.0 $\mu_B$ per Ni. Even though we have presented our results for La, the same type of solution can be derived for the NdNiO$_2$ counterpart (see Appendix \ref{appendix_C}, Fig. \ref{fig9}).

In order to illustrate the lack of La-$d$ - Ni-$d$ hybridization in this case, we have highlighted in the band structure of Fig. \ref{fig4} the Ni-$d_{z^2}$ character for both Ni atoms in the structure. These $d_{z^2}$ bands become completely occupied and are very flat (in contrast to the bulk plot of Fig. \ref{fig3}) and no Ni-$d_{z^2}$ character can be seen in the unoccupied part of the spectrum. The DOS plot helps locate the different characters of other bands shown. La-$d$ bands start appearing above 1 eV (only one spin channel is analyzed for clarity). The O-$p$ bands start to emerge below -3 eV in this case, further away from the Fermi level than in the cuprates \cite{botana_2020_similarities,botana2021low}.

\begin{figure}[ht]
\begin{center}
\includegraphics[width=\columnwidth,draft=false]{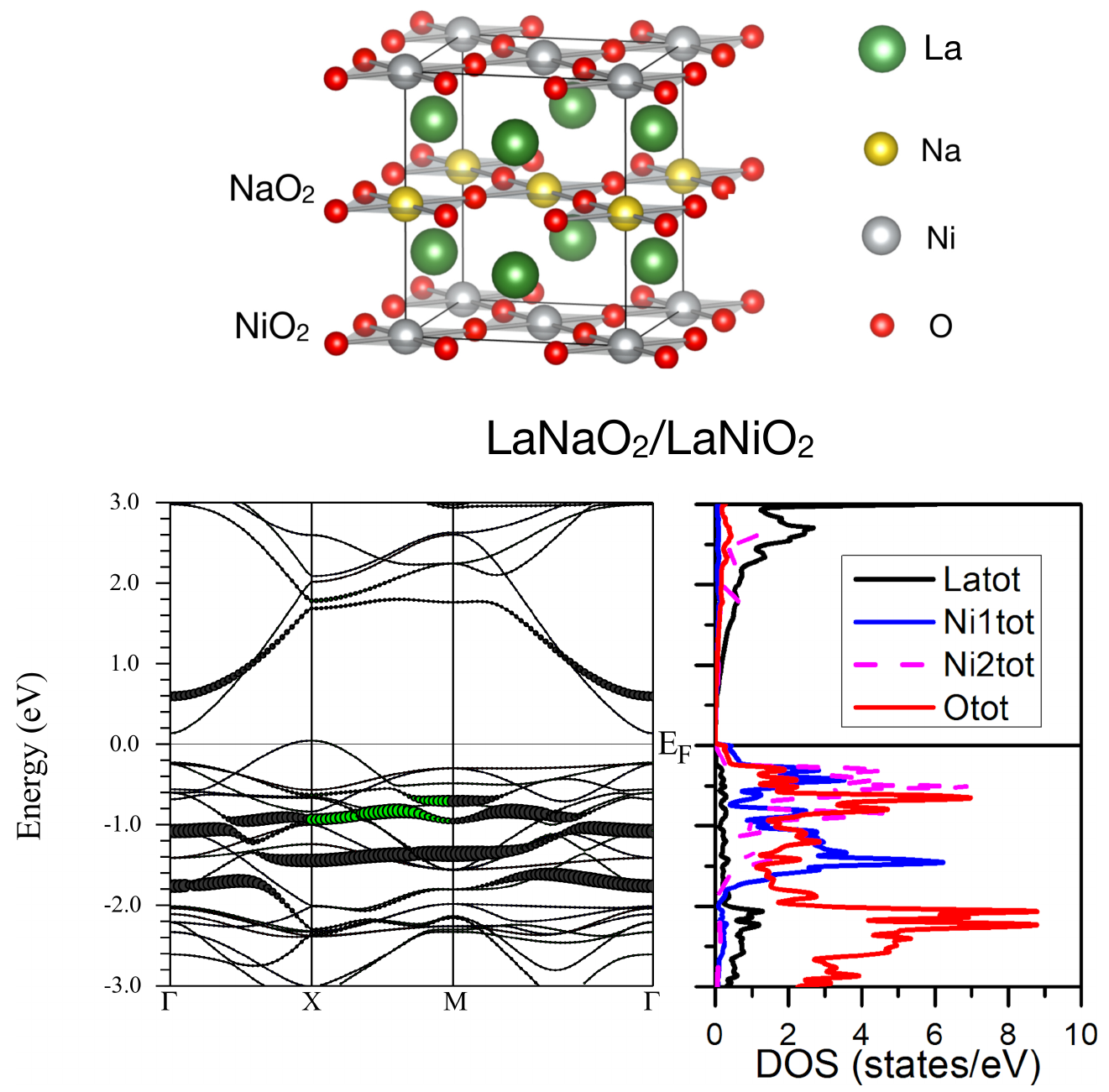}
\caption{Top panel. Supercell constructed alternating one layer of LaNiO$_2$ and one layer of LaNaO$_2$. The Ni-Ni direct off-plane hopping is blocked structurally by a single layer of LaNaO$_2$. However, the La-$d$ off-plane hopping is not blocked. Na atoms in yellow, La atoms in green, O atoms in red, Ni atoms in gray. Bottom left panel. Band structure of a multilayer (LaNaO$_2$)$_1$/(LaNiO$_2$)$_1$ obtained with a very small $U$= 1.5 eV. A gap opening together with AF ordering occurs when the off-plane coupling is geometrically suppressed, even when the La-$d$-La-$d$ off-plane hopping is permitted. Only one spin channel is shown and the La-$f$ states are unoccupied and away from the Fermi level. The Ni-$d_{z^2}$ bands are highlighted for both Ni atoms antiferromagnetically coupled (a Hund splitting can be noticed between them as before). Bottom right panel. Corresponding La, O (only the O atoms in the NiO$_2$ plane are shown), and Ni atom-resolved density of states.}
\label{fig5}
\end{center}
\end{figure}

In order to study the band(s) that bring about metallicity in the bulk in more detail, we have constructed a hypothetical system, like the one shown in Fig. \ref{fig5}. This is a multilayer  that alternates one LaNiO$_2$ and one LaNaO$_2$ unit cell stacked along the (001) direction. This system provides continuity along the $c$-axis for the rare-earth La-$d$-La-$d$ direct hopping. However, in this structure the hopping paths for the transition metal Ni-$d$ bands are largely confined into the $ab$ plane. Their off-plane hopping is largely blocked by the LaNaO$_2$ layer, since the Na atoms do not provide  continuity for a direct off-plane Ni-$d_{z^2}$  hopping. Also,  the La-$d$-Ni-$d$ hybridization  is largely hampered by the NaO$_2$ blocking layer. Note that LaNaO$_2$ exists but it is not isostructural to LaNiO$_2$ \cite{lanao2_structure}. Here, for the purpose of comparison though, we assume the same crystal structures taking the in-plane lattice parameter of LaNiO$_2$ as a basis, and then all the internal atomic positions and the off-plane lattice parameter have been relaxed. It is not relevant for our discussion whether this system is experimentally achievable or not, it just provides another way of suppressing the $c$-axis dispersion to confirm the reasoning exposed above. We note that the equivalent Li-based system (LiLa$_2$NiO$_4$) has been shown to be dynamically stable \cite{arita_design}.

The band structure of the multilayer is shown in Fig. \ref{fig5}.  
Similar to the LaNiO$_2$ monolayer, the system is an AF insulator. The only subtle difference is that a very small $U$ ($\sim$ 1.5 eV) is needed for a gap to be opened up (an even smaller $U$ is required in the case of NdNiO$_2$, see Appendix \ref{appendix_C}, Fig \ref{fig10}). This gapped AF solution occurs even though there is a continuity of the rare-earth sublattice along the off-plane direction. This confirms that the key factor in order to open up a gap is for the geometry to provide a way to suppress the R-Ni off-plane hopping that leads to hybrid bands with a large dispersion, crossing the Fermi level. 
When the Ni-$d_{z^2}$ bands (highlighted in the band structure shown) have no mechanism to hybridize along the $c$-direction, these bands remain very flat and fully occupied. Then, the R-$d$ bands do not disperse below the Fermi level and the system is an analog of the reduced RP phases and parent cuprates at $d^9$ filling: an AF insulator, with no self-doping. As such, at the electronic structure level, this computational experiment of a multilayered system works similar to the single-layer LaNiO$_2$. All in all, we can conclude that reducing dimensionality is a way to reach a parent phase that is AF and insulating in parent RNiO$_2$, like that in the cuprates.

It is important to explain this result in detail. If one looks at the band character of the band crossing the Fermi level in bulk LaNiO$_2$ \cite{KWLee2004_prb} or NdNiO$_2$ \cite{Choi2020}, it is mostly R-$d$ in character. If this were the whole story, in this multilayered system, such a band would still have room to be largely dispersive. What our calculations show is that its large dispersion requires a strong hybridization with the Ni-$d_{z^2}$ band. When this hybridization is cut by, e.g. introducing NaO$_2$ planes in the structure, then this band is no longer living close to the Fermi level and the electronic structure becomes cuprate-like, AF correlations become stronger (larger moments closer to the nominal ionic $S=1/2$ value), and the ground state becomes AF and insulating even at the GGA level (or with a very small $U$ in the case of the multilayer).

\section{discussion}

We have shown that an AF insulating ground state can be obtained in R$_{n+1}$Ni$_n$O$_{2n+2}$ layered nickelates at $d^9$ filling: in a natural manner in the reduced RP phases ($n=3-6$), or introducing a structural element to cut the $c$-axis dispersion in the $n=\infty$ compounds.  

In the infinite-layer nickelates (RNiO$_2$) we have shown that the origin of the celeb self-doping effect is not simply in the rare-earth $d$ bands -- the existence of the electron pockets with R-$d$ parentage requires a strong hybridization with the Ni-$d_{z^2}$ bands. This hybridization leads to the formation of wide bands that arise due to the large hopping taking place along the $c$-axis. This provides further evidence of the active role of the Ni-$d_{z^2}$ orbitals in the electronic structure of infinite-layer nickelates \cite{choi2020fluctuation, lechermann20prx, werner20Hund, kotliar20prb}. 

This situation (a highly dispersive band crossing the Fermi level) does not occur in the parent cuprates. There are various reasons for this. In the isostructural infinite-layer cuprate (CaCuO$_2$) the Ca-$d$ bands are very far above the Fermi level to be able to produce any effects in the electronic structure. In other cuprates containing atoms with available unoccupied $d$ bands (such as the La bands in La$_2$CuO$_4$ or the Y bands in the YBCO family), there are important structural differences. The case of La$_2$CuO$_4$ allows for a direct comparison with the layered nickelate family. Such a structure \cite{la2cuo4_structure} consists of CuO$_2$ planes but separated by a spacing La$_2$O$_2$ blocking layer (the same can be said about YBCO). These structural elements provide a barrier for the off-plane hoppings  (in a similar fashion to the fluorite slab in finite-$n$ members of the layered nickelate family, and the NdNaO$_2$ layer in the hypothetical multilayered system we described for the infinite-layer nickelate) ultimately allowing for an AF insulating phase to appear naturally in the $d^9$ limit.

To summarize, we have shown that a parent insulating AF phase can take place in the phase diagram of layered  nickelates (R$_{n+1}$Ni$_{n}$O$_{2n+2}$) close to half-filling. 
Our work shows yet another piece of evidence of the similarities between layered nickelates and cuprates, which were not apparent at first but can be elucidated digging deeper into the properties of these new nickel-oxide materials.

\section{Acknowledgments}
V.P. is supported by the Ministry of Science of Spain through the project PGC2018-101334-B-C21.  H.L. and A.S.B acknowledge NSF Grant No. DMR 2045826. We acknowledge the ASU Research Computing Center for HPC resources.


%


\newpage


\appendix

\section{Density of states of higher-ordered nickelates}\label{appendix_A}

\begin{figure*}[tb]
    \centering
    \includegraphics[width=15.0cm]{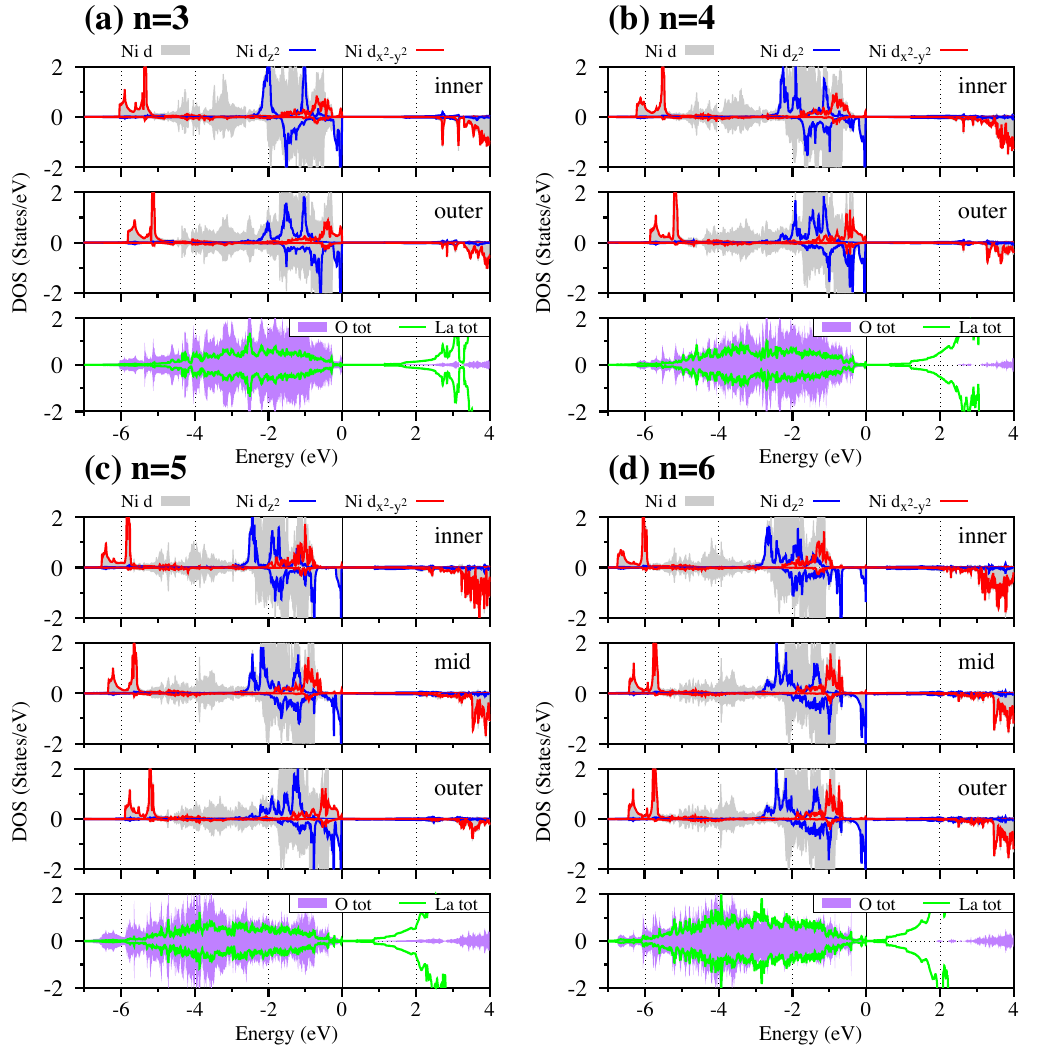}
    \caption{Orbital-projected density of states of higher-order layered RP phases in GGA+$U$ ($U$= 5 eV, $J$= 0.7 eV) at $d^9$ filling }
    \label{fig6}
\end{figure*}

The projected densities of states (PDOS) of La-based $n$=3-6 reduced nickelates corresponding to the band structures in Fig. \ref{fig2} are shown in Fig. \ref{fig6}. For all materials, the $d_{z^2}$ orbitals, which are dominantly located from -2 eV all the way up to the Fermi energy, are fully occupied. The only Ni-$d$ band that remains unoccupied is a single
minority-spin $d_{x^2-y^2}$ band per Ni. 
The oxygen-$p$ states (shown in Fig. \ref{fig6} in purple color) appear at lower energies. These O-$p$ states shift up in energy with decreasing $n$.

\section{Fat bands showing the $d_{x^2-y^2}$ character}
\label{appendix_B}

\begin{figure}[h!]
\centering
\begin{tabular}{ccc}
\begin{minipage}{.47\textwidth}
  \centering
  \includegraphics[width=\columnwidth,draft=false]{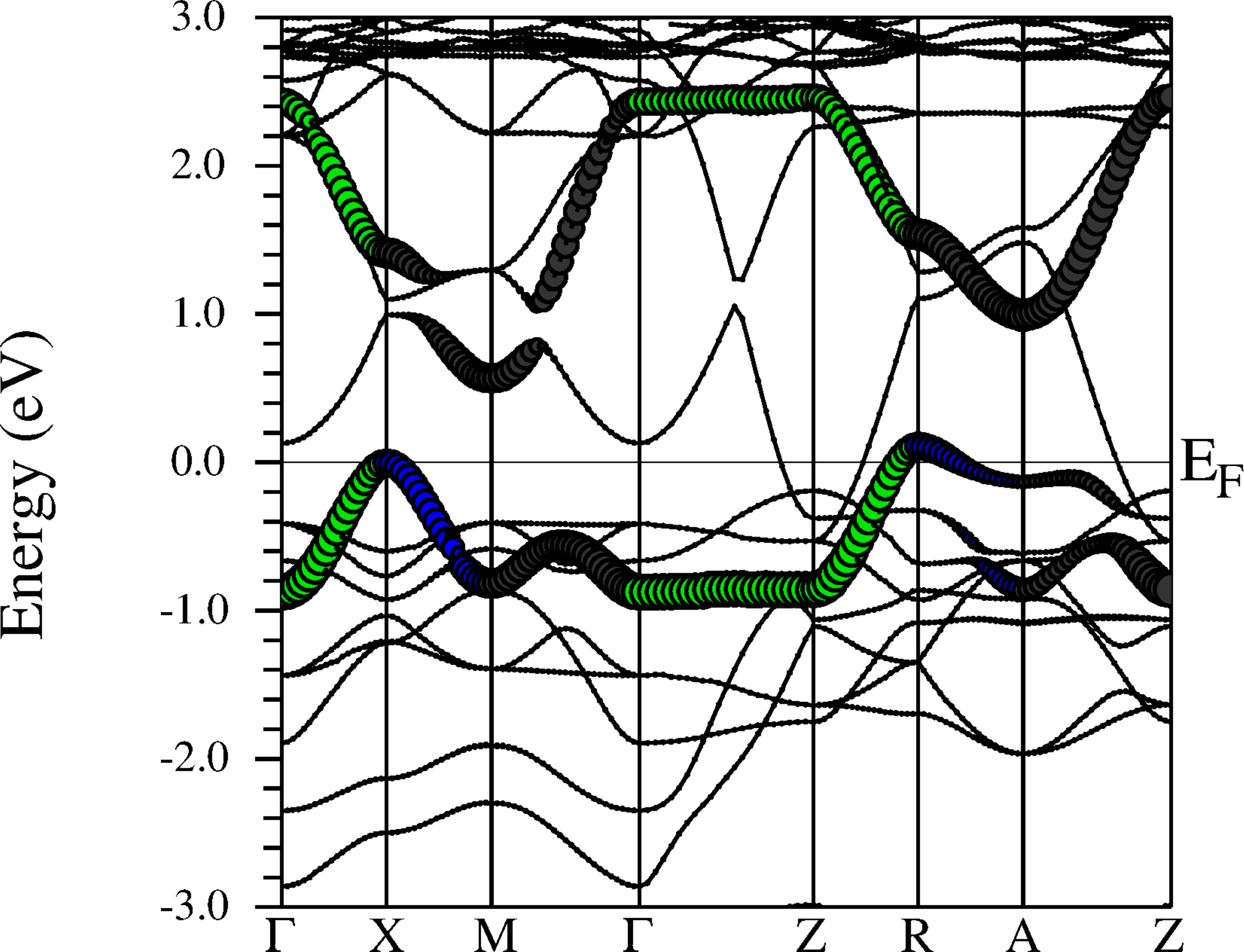}
    \caption{Band structure of bulk LaNiO$_2$ in C-type AF configuration (in-plane $\sqrt{2} \times \sqrt{2}$ unit cell). The bands highlighted correspond to the $d_{x^2-y^2}$ fat bands coming from the two inequivalent Ni atoms in the structure (exchange split due to the AF coupling). For each spin channel, there is one unoccupied $d_{x^2-y^2}$ band and another one which becomes partly depopulated due to the self-doping effect.}
    \label{fig7}
\end{minipage}%
&~&
\begin{minipage}{.47\textwidth}
  \centering
  \includegraphics[width=\columnwidth]{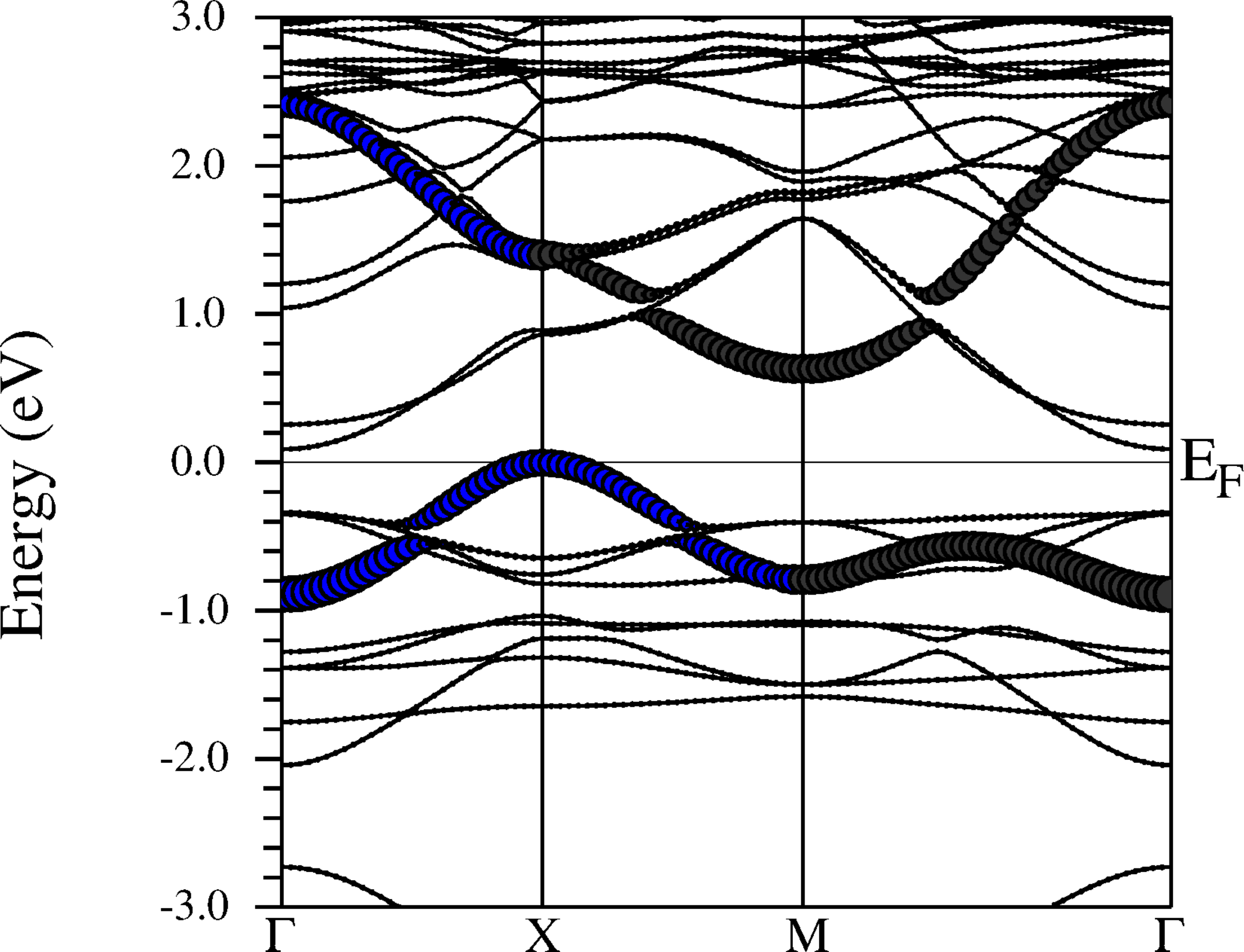}
  \caption{Band structure of monolayer LaNiO$_2$ in the checkerboard AF configuration (in-plane $\sqrt{2} \times \sqrt{2}$ unit cell). The bands highlighted correspond to the $d_{x^2-y^2}$ fat bands coming from the two inequivalent Ni atoms in the structure (exchange split due to the AF coupling). We see that for each spin channel (only one shown) there is a fully unoccupied $d_{x^2-y^2}$ band and another one that is fully occupied, leading to a fully gapped electronic structure.}
    \label{fig8}
\end{minipage}
\end{tabular}
\end{figure}

We provide in Figs. \ref{fig7} and \ref{fig8} the fat bands for bulk and monolayer LaNiO$_2$ highlighting the $d_{x^2-y^2}$ orbital contribution, in order to complement the corresponding $d_{z^2}$ bands discussed in the main text. In these plots, we can observe that the self-doping effect in the bulk produces a partial depopulation of the Ni $d_{x^2-y^2}$ band. However, in the monolayer limit, the self-doping effect is no longer present and the $d_{x^2-y^2}$ is completely half-filled. This latter argument applies to the multilayer system as well. 

\section{NdNiO$_2$ systems}
\label{appendix_C}

\begin{figure}[h!]
\centering
\begin{tabular}{ccc}
\begin{minipage}{.47\textwidth}
  \centering
  \includegraphics[width=\columnwidth]{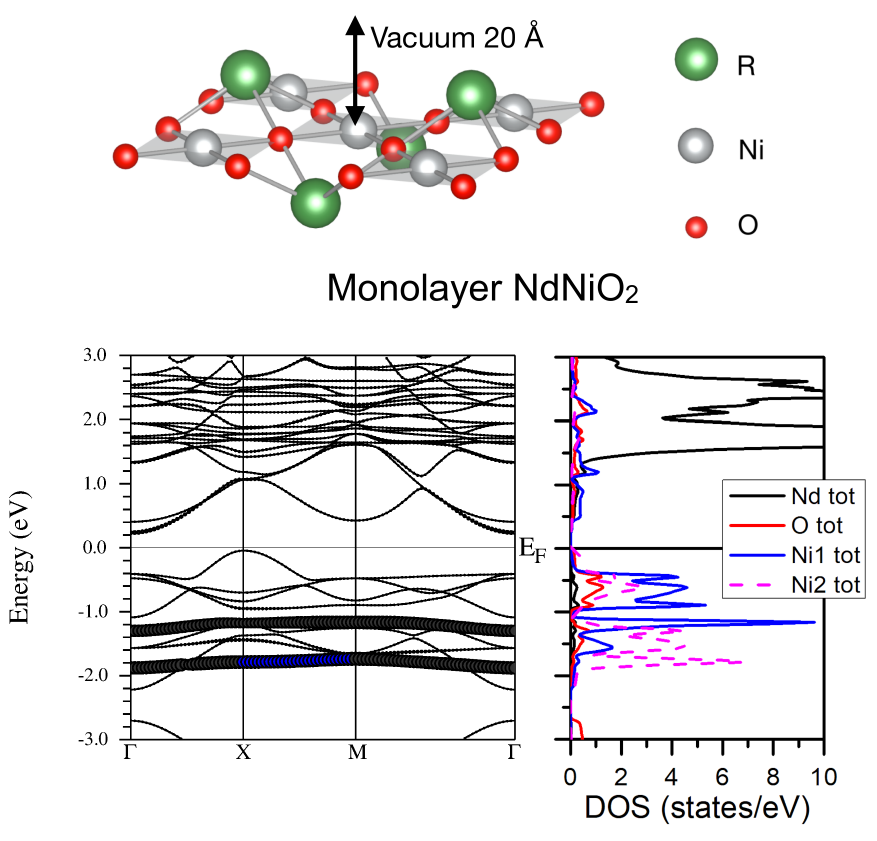}
    \caption{Top panel. Structure of a  NdNiO$_2$  monolayer. Bottom left panel. Band structure of the NdNiO$_2$ monolayer with AF ordering of the Ni atoms, showing only the minority spin channel (Nd moments are parallel). The Ni-$d_{z^2}$  states are highlighted for both Ni atoms (a Hund splitting can again be observed between them). The flat Nd-$f$ bands (unoccupied) appear 2-3 eV above the Fermi level. The O-$p$ bands (not shown) start to appear below -3 eV. A gap opening together with the antiferromagnetic ordering occurs when the off-plane coupling is suppressed. This is visible even at the $U= 0$ limit (bands shown), similar to the situation in bulk CaCuO$_2$, and unlike the self-doped metallic phase appearing in bulk NdNiO$_2$. Bottom right panel. Corresponding Nd, O and  Ni atom-resolved density of states (both Ni atoms have opposite spins). }
    \label{fig9}
\end{minipage}%
&~&
\begin{minipage}{.47\textwidth}
  \centering
  \includegraphics[width=\columnwidth,draft=false]{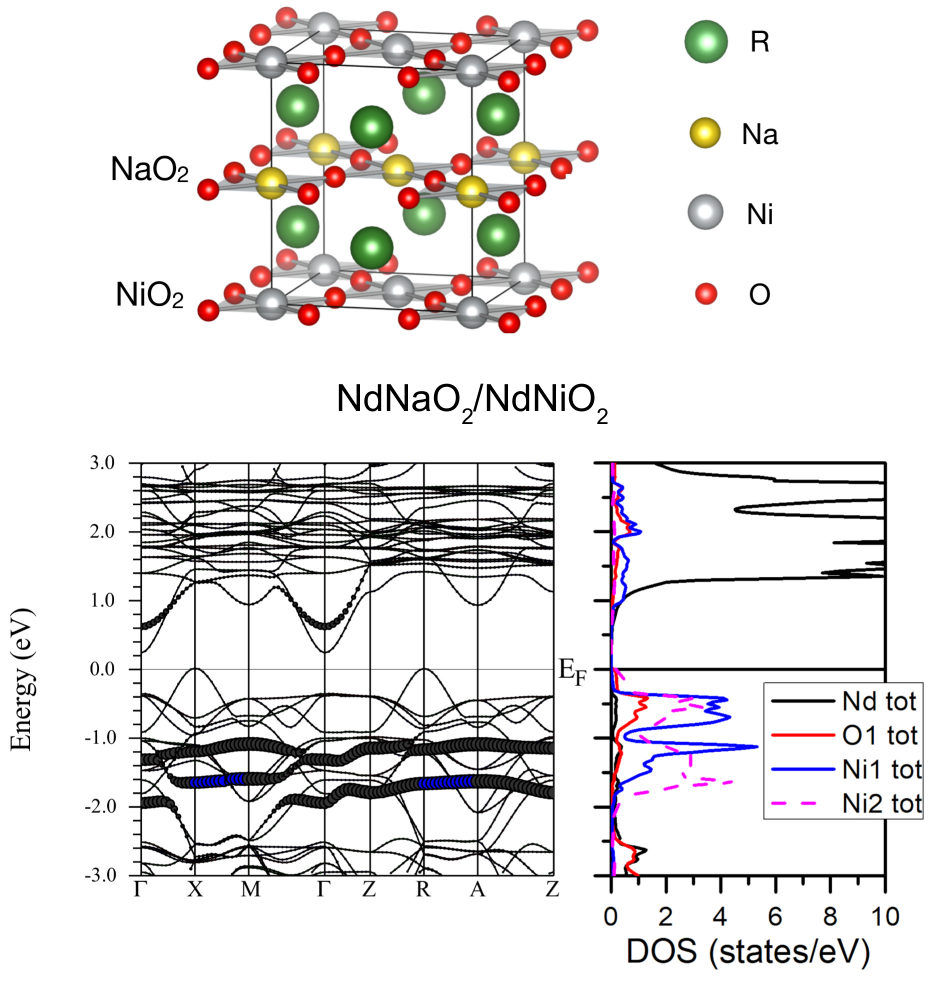}
    \caption{Top panel. Supercell constructed alternating one layer of NdNiO$_2$ and one layer of NdNaO$_2$.  The Ni-Ni direct off-plane hopping is blocked structurally by a single layer of NdNaO$_2$. However, the Nd-$d$ off-plane hopping is not blocked. Na atoms in yellow, R atoms in green, O atoms in red, Ni atoms in gray. Bottom left panel. Band structure of a multilayer (NdNaO$_2$)$_1$/(NdNiO$_2$)$_1$ obtained with a very small $U$= 0.7 eV. A gap opening together with antiferromagnetic ordering occurs when the off-plane coupling is geometrically suppressed, even when the Nd-$d$-Nd-$d$ off-plane hopping is permitted. Only the minority spin is shown so that the Nd-$f$ states are unoccupied and away from the Fermi level. The Ni-$d_{z^2}$ bands are highlighted for both Ni atoms antiferromagnetically coupled (a Hund splitting can be noticed between them, as before). Bottom right panel. Corresponding Nd, O (only the O atoms in the NiO$_2$ plane are shown), and Ni atom-resolved density of states.}
    \label{fig10}
\end{minipage}
\end{tabular}
\end{figure}

We provide a description of the electronic structure of the NdNiO$_2$ systems analogous to the La-based ones presented in the main text (with the same computational parameters described above). Figure \ref{fig9} shows the band structure and DOS of the minority-spin channel of the NdNiO$_2$ monolayer (the Nd moments are considered all parallel so that in the minority-spin channel, the Nd-$f$ bands are completely unoccupied away from the Fermi level). The Ni $d_{z^2}$ bands are completely occupied and present a small dispersion.

Fig. \ref{fig10} shows a multilayer system analogous to that in the main text, but using Nd-based materials. Again, the presence of a blocking structure that cuts the $c$-axis dispersion leads to a gap opening at the Fermi level. In this case, an even smaller $U$ is able to open the gap when compared to the La-based case. Again, the electronic structure is presented for the minority-spin channel only, with the Nd-$f$ bands unoccupied well above the Fermi level.

\end{document}